\documentclass[preprint,showpacs,amsmath,amssymb,aps,prd,nofootinbib]{revtex4}
\usepackage{epsfig}

\usepackage{xcolor}
\usepackage{epstopdf}

\begin{document}

\def\bea{\begin{eqnarray}}
\def\eea{\end{eqnarray}}
\def\nn{\nonumber}
\newcommand{\snu}{\tilde \nu}
\newcommand{\sll}{\tilde{l}}
\newcommand{\asnu}{\bar{\tilde \nu}}
\newcommand{\stau}{\tilde \tau}
\newcommand{\dmsnu}{{\mbox{$\Delta m_{\tilde \nu}$}}}
\newcommand{\mt}{{\mbox{$\tilde m$}}}

\renewcommand\epsilon{\varepsilon}
\def\be{\begin{eqnarray}}
\def\ee{\end{eqnarray}}
\def\lla{\left\langle}
\def\rra{\right\rangle}
\def\za{\alpha}
\def\zb{\beta}
\def\lsim{\mathrel{\raise.3ex\hbox{$<$\kern-.75em\lower1ex\hbox{$\sim$}}} }
\def\gsim{\mathrel{\raise.3ex\hbox{$>$\kern-.75em\lower1ex\hbox{$\sim$}}} }
\newcommand{\Rbs}{\mbox{${{\scriptstyle \not}{\scriptscriptstyle R}}$}}

\setlength\oddsidemargin{0cm}
\setlength\evensidemargin{0cm}

\setlength\textwidth{16cm}
\setlength\textheight{22.5cm}
\setlength\topmargin{-1.0cm}

\draft
%\preprint{{\vbox\hbox{}
%\hbox{May 2014} }}}

%@twocolumnfalse\endcsname

\title{ keV Sterile Neutrino Dark Matter and Low Scale Leptogenesis}

%\renewcommand{\theequation}{\arabic{section}.\arabic{equation}}
%\newcommand{\Slash}[1]{\ooalign{\hfil/\hfil\crcr$#1$}}
%\newcommand{\\}{cr}
%\newcommand{\zfn}{\renewcommand{\thefootnote}{\fnsymbol{footnote}}}
%\newcommand{\Zfn}{\zfn\footnote}
%#!platex
%\topmargin=0cm
%\oddsidemargin=17.0cm \evensidemargin=0truecm
%\textheight=17.0cm \textwidth=11.0cm
%\textwidth 16.5cm
%\textheight 20.5cm
%\oddsidemargin 0.3cm

\thispagestyle{empty}
%\begin{titlepage}
\author{ Sin Kyu Kang\footnote{ E-mail: skkang@snut.ac.kr} and  Ayon Patra\footnote{E-mail: ayon@okstate.edu}}
%\date{\today}
\affiliation{ Institute for Convergence Fundamental Study, Seoul National University of Science and Technology, Seoul 139-743, Korea }

\begin{abstract}
We consider a simple extension of the Standard Model to consistently explain the observation of a peak in the galactic X-ray spectrum at 3.55 keV and the light neutrino masses along with the baryon asymmetry of the universe. 
The baryon asymmetry is generated through leptogenesis, the lepton asymmetry being generated by the decay of a heavy neutrino with TeV mass scale. The extra singlet fermion introduced in the model can be identified as a dark matter candidate of mass 7.1 keV. It decays with a lifetime much larger than the age of the universe, producing a final state photon. 
The Yukawa interactions between the extra singlet neutrino and a heavier right-handed neutrino play a crucial role in simultaneously achieving low scale leptogenesis and relic density of the keV dark matter candidate.

\end{abstract}
\pacs{ 98.80.-k, 95.35.+d, 14.60.St, 14.60.Pq }
 \maketitle \thispagestyle{empty}
% \narrowtext
%
%\widetext
%\clearpage
%%%%%%%%%%%%%%%%%%%%%%%%%%%%%%%%%%%%%%%%%%%%%%%%%%%%%%%%%%%%%%%%%%%%%%%%%%%%
%%%
%\tableofcontents
%\clearpage
%%%%%%%%%%%%%%%%%%%%%%%%%%%%%%%%%%%%%%%%%%%%%%%%%%%%%%%%%%%%%%%%%%%%%%%%%%%%
%%%
%\textwidth 18cm
%\textheight 22.5cm

\section{Introduction}

The identity of the dark matter remains a mystery in particle physics and cosmology.
Recent observation of a peak in the galactic X-ray spectrum at 3.55 keV \cite{x1,x2} can be interpreted as a signal of the existence of a 7.1 keV dark matter candidate. This dark matter can decay to produce final state photons which can explain the observed X-ray signal. 
%
%
%Recently, two independent groups have reported the observation of a peak in the galatic X-ray spectrum
%at 3.55 keV \cite{x1,x2}, which can not be explained in terms of known physics and astrophysics.
%If confirmed, this discovery will have a major impact in astrophysics, cosmology and particle physics.
%A natural interpretation of this result is in terms of the decay of a new particle with mass of
%7.1 keV into photons.
%This new particle should be around today for its decay to be observed, which naturally suggests that
%the decaying particle is also the dark matter of the Universe. 
%Such a warm dark matter particle in this mass range \cite{warmDM} has been known to fit
%other cosmological observations quite well, in some instances even better than the canonical
%cold dark matter 
A variety of self-consistent models \cite{DM-model} have been studied to explain the
cosmological data as well as the galactic X-ray spectrum and confront them with experiments.

Sterile neutrinos can not only be a good candidate for warm dark matter \cite{warmDM1,warmDM2}, but also play an essential role in achieving smallness of neutrino masses \cite{seesaw} and baryogenesis via leptogenesis \cite{lepto}. The sterile neutrino states can mix with the active neutrinos and such admixtures contribute
to various processes which are forbidden in the Standard Model and can affect the interpretations of cosmological
and astrophysical observations \cite{sterilen}. As required from the observation of the galactic X-ray spectrum, they can radiatively produce photons
through their mixing with active neutrinos \cite{pal}.
Thus, the masses of the sterile neutrinos and their mixing with
the active neutrinos are subject to current experimental, cosmological and astrophysical constraints as well as 
the recent observation of the galactic X-ray spectrum.

In ref.\cite{skk1}, a model for low scale leptogenesis has been proposed.
In the model, tiny neutrino masses are achieved  by introducing
extra singlet neutrino on top of heavy right-handed Majorana neutrinos. 
The extra Yukawa couplings between singlet neutrinos allow us to lower the scale of leptogenesis down to TeV scale.
In this paper, we revisit this model to investigate whether the extra singlet neutrino with 7.1 keV mass can be a dark matter leading to 3.55 keV photon peak. We will show how the required size of the  mixing parameter between the extra sterile neutrino and active neutrinos can be achieved in the model so as to explain the galactic X-ray spectrum.  Since the extra singlet neutrino in the model is out of equilibrium for any time after inflation, 
it must be produced by non-thermal mechanism.
We will show that the light sterile neutrino can be produced via the out-of-equilibrium decay of singlet scalar field \cite{nonthermal}
which is initially assumed to be in thermal equilibrium in early Universe and then decoupled at the temperature around the scale of singlet scalar mass of order 100 GeV.
We do not consider the possible production mechanism of the warm dark matter candidate via neutrino mixing 
at low temperature \cite{warmDM1}
because it is not dominant  for the sterile neutrino with mass above 3.5 keV \cite{nonthermal}.
 As will be shown later, the coupling between
the extra singlet neutrino and singlet scalar plays an important role in achieving low scale leptogenesis and the
right amount of relic density for the sterile neutrino as a dark matter.
So in this model, there will be a connection between lepton asymmetry and relic density of dark matter.   

In sec. II, we set up the model and show how the extra sterile neutrino has mass of around 7.1 keV
and that we can get the required size of the mixing angle between the extra sterile neutrino and active neutrinos
so as to accommodate the result of the galactic X-ray spectrum.
In sec. III, we show how low scale leptogenesis can be realized in this model.
Sec. IV is devoted to the derivation of the relic density of the extra sterile neutrino as a dark matter.
Here, we will discuss how the lepton asymmetry can be connected to the relic density of the extra singlet neutrino. In Sec. V we look at the possible decay channels of the sterile neutrino dark matter candidate and the production of 3.55 keV photon signal. Some comments and conclusion will be given in Sec. VI. 

\section{Model and Dark Matter Candidate}
On top of the seesaw model, we introduce a singlet neutrino, $\chi$, and a singlet scalar field, $\eta$.
The mass terms and Yukawa interactions in the Lagrangian we consider is given in the charged
lepton basis as \cite{skk1}
\be
{\cal L}=M_{N_i}N_i^T N_i+Y_{D_{ij}} \bar{\L}_i
H  N_j+ Y_{\chi_{i}} \bar{\chi}\eta N_i -\mu \chi^T \chi +h.c.~,
\ee
where $\L_i,N_i, H $ stand for SU(2)$_L$ lepton doublet with flavor index $i$, right-handed singlet neutrino and an SU(2)$_L$ scalar doublet field respectively.
We impose a $Z_2$ symmetry under which the fields $\eta$ and $\chi$ are odd and all the other fields
are even.
This $Z_2$ symmetry, if unbroken, would immediately imply that the lightest singlet neutrino
$\chi$ is stable and can be a dark matter candidate. This is not strictly true for our model as the $Z_2$ symmetry is broken once the scalar singlet $\eta$ gets a non-zero vacuum expectation value. Although $\chi$ is no longer stable, it is still a viable dark matter candidate since its lifetime is much longer than the age of the universe as will be discussed later. 

The scalar potential in this model is given as \cite{scalar}
\be
{\cal {L}}_{Scalar}=\frac{\mu_H^2}{2} H^{\dagger} H+\frac{\lambda_H}{4}(H^{\dagger}H)^2+\frac{\mu_\eta^2}{2} \eta^2 +\frac{\lambda_\eta}{4}\eta^4+\frac{\lambda_{H \eta}}{2}(H^{\dagger}H)\eta^2.
\ee
Defining the vacuum expectation values of the neutral component of $H$ and $\eta$ fields as $v$ and $u$ respectively, we expand the fields as $H \rightarrow \frac{h}{\sqrt{2}}+v$ and $\eta \rightarrow \eta+u$. This gives us a $2\times2$ mass squared matrix for the scalar fields given as
\be
\begin{pmatrix} \lambda_H v^2&\sqrt{2}\lambda_{H\eta}uv\\\sqrt{2}\lambda_{H\eta}uv&2\lambda_\eta u^2 \end{pmatrix}.
\ee
Assuming $\lambda_{H\eta} << \lambda_H,\lambda_\eta$ we get the eigenvalues of this mass squared matrix as
\begin{eqnarray}
M_H^2 &=& \lambda_H v^2+\frac{\lambda_{H\eta}uv}{\sqrt{2}}\tan{2\theta}, \notag \\
M_\eta^2 &=& 2\lambda_\eta u^2-\frac{\lambda_{H\eta}uv}{\sqrt{2}}\tan{2\theta},
\label{eq:scalar}
\end{eqnarray}
where $\theta$ is the scalar mixing angle given as 
\be
\tan{2\theta}=\frac{2\sqrt{2}\lambda_{H\eta}uv}{\lambda_Hv^2-2\lambda_\eta u^2}. 
\ee

For $\eta$ to be an appropriate candidate to act as the heat bath we should have $\lambda_{H\eta} > 10^{-6}$ \cite{nonthermal}. On the other hand we need to make sure that its mixing with the Higgs boson is small enough to not effect the Higgs signals. We thus choose $\lambda_{H \eta} \sim 10^{-5}$. For such a small $\lambda_{H\eta}$, we can easily neglect its contribution to the masses of the scalars given in Eq.~({\ref{eq:scalar}}). The mass of the Standard Model like Higgs boson is $M_H=125$ GeV for $v=174$ GeV and $\lambda_H=0.516$. We choose the vacuum expectation value of the scalar singlet $u=100$ GeV, and $\lambda_\eta=0.254$ which gives $M_\eta=71.3$ GeV as a benchmark point for achieving
right amount of relic density for the dark matter candidate.

Let us now focus our attention to the neutrino sector in this model. The neutrino mass matrix in the basis $(\nu_j,N_i,\chi)$ is given as \cite{skk1}
\be
M_{\nu}=\left(\begin{array}{ccc}
 0 & m_{D_{ij}} & 0 \\
 m_{D_{ij}} & M_{N_{ii}} & M_{\chi_{i}} \\
 0 & M_{\chi_{i}} & -\mu \end{array}\right), \label{massmatrix}
\ee
where $m_{D_{ij}}=Y_{D_{ij}}<H>, M_{\chi_{i}}=Y_{\chi_{i}}<\eta>$.
Here we assume that $M_N \gg M_{\chi} \gg \mu, m_{D}$. 
After integrating out the right-handed heavy neutrino sector $N$ in the above
Lagrangian, 
and block diagonalization of the effective mass terms,
the light neutrino mass matrix and mixing between the
active and sterile neutrinos are given by
\be
m_{\nu} &\simeq & \frac{1}{2}\frac{m_D}{M_{\chi}}~\mu~ \left(\frac{m_D}{M_{\chi}}\right)^T,
                  \label{dsw}\\
\tan2\theta_{\chi} &=& \frac{2m_D M_\chi}{M^2_\chi+4\mu M_N-m_D^2}~, \label{mixing}
\ee
where we omitted the indices of the mass matrices, $m_D,M_\chi,M_N$ and $\mu$
for simplicity.

%We note that the term $m_D^2/M_N$ corresponding to
%typical seesaw (type I) mass is cancelled out.
On the other hand, the sterile neutrino mass is approximately
given as
\begin{eqnarray}
m_{\chi} \simeq \mu + \frac{M^2_\chi}{4M_N}. \label{sterile}
\end{eqnarray}
We further assume that $M^2_\chi \ll 4 \mu M_N$ and then, the mixing angles
$\theta_\chi$ and the sterile neutrino mass $m_{\chi}$ are approximately
given by
\be
\tan2\theta_{\chi} &\simeq & \sin2\theta_{\chi}  \simeq 
  \frac{m_D M_{\chi}}{2\mu M_N},  \label{mixing}\\
m_{\chi} &\simeq &\mu
 \label{sterile2}
\ee
%We can see from Eq. (\ref{sterile2}) that is mainly responsible for the value of $m_{\chi}$.
Taking $\mu \sim 7.1$ keV, the sterile neutrino is regarded as dark matter leading to 3.55 keV
peak of the galactic X-ray spectrum.
Then, we see from Eq.(\ref{dsw}) that $m_D/M_\chi \simeq 3.8 (1.7)\times 10^{-3}$ for $m_{\nu}\simeq 0.05 (0.01)$ eV corresponding to the atmospheric (solar) neutrino mass scale in the hierarchical spectrum.
To achieve low scale leptogenesis, it is desirable to take the lightest $M_N$ to be a few TeV.
Once we take $M_N\sim 10$ TeV,  we are led from Eq.(\ref{mixing}) to
\be
\theta_{\chi}  \simeq \frac{m_D M_\chi}{4 \mu M_N}\sim \left(\frac{M_\chi}{0.86 (1.3)\times 10^{10} \mbox{eV}}\right)^2. \label{s-mixing}
\ee
In order for the sterile neutrino to be the dark matter with mass $7.1$ keV, it is required 
$\sin2\theta_{\chi}\sim 10^{-5}$ \cite{x1,x2}.
Combining this constraint with Eq.(\ref{s-mixing}), the scale of $M_{\chi}$ is determined to be
around $ 20(30)$ MeV, and thus  $m_{D}$ to be 70(50) keV.
Those scales of $m_{D}$  and  $M_{\chi}$ are achieved by taking $Y_{D}\sim 10^{-6}$ and
$Y_{\chi}\sim 10^{-4}$, respectively for $<H>\simeq 246$ GeV and $<\eta>\sim 100$ GeV.
When we take $M_N$ to be less 10 TeV, then we get smaller $M_\chi$ and $m_D$.

We note that such a small mixing angle $\theta_\chi$ between sterile and active neutrinos ensures that
sterile neutrinos were never in thermal equilibrium in the early
Universe and this allows their abundance to be smaller than the predictions in thermal equilibrium.
Thus, the right abundance of the sterile neutrino should be achieved not by thermal production frozen out
below sterile neutrino mass
but production via freeze-in decay of the scalar $\eta$ as will be shown in section \ref{dmp}.
% Moreover, a sterile neutrino with these
%parameters is important for the physics of supernova, which can
%explain the pulsa kick velocities \cite{pulsa}.
% In addition, there are some bounds on the mass of sterile neutrino from the
%possibility to observe sterile neutrino radiative decays from
%X-ray observations and Lyman $\alpha-$forest observations of order of a few keV.

\vspace{0.3cm}
\section{Leptogenesis}
Now, let us consider how low scale leptogenesis can be
successfully achieved in our scenario by the
decay of the lightest right-handed Majorana neutrino before the
scalar fields get vacuum expectation values.
We take the basis where the mass terms $M_{N_{ij}}$ and $\mu$ are real and
diagonal. In this basis, the elements of $Y_D$ and $Y_{\chi}$ are in
general  complex. The lepton number asymmetry required for
baryogenesis is given by
\begin{eqnarray}
\epsilon_{1} &=& -\sum_i\left[\frac{\Gamma(N_1
\to \bar{l_i}H^{\ast}) - \Gamma(N_1 \to l_i H) }{\Gamma_{\rm
tot}(N_1)}\right] ,
\end{eqnarray}
where $N_1$ is the lightest right-handed neutrino and $\Gamma_{\rm tot}(N_1)$
is the total decay rate given to leading order  by
\begin{equation}
  \label{eq:vv}
\Gamma_{\rm tot}(N_1)={(Y_D^\dagger Y_D)_{11}+|Y_{\chi_1}|^2
\over 4\pi}M_{N_1},
\end{equation}
where we assume that the masses of the Higgs sectors
and extra singlet neutrinos are much smaller compared to $M_{N_1}$.
 In addition to the diagrams of the standard
leptogensis scenario \cite{Covi}, there is a new contribution of the
diagram which corresponds to the self energy correction of the
vertex arisen due to the Yukawa interaction $Y_{\chi} \bar{\chi}\eta N$,
so that the lepton number asymmetry is given by
\begin{equation}
\epsilon_1 = \frac{1}{8\pi} \sum_{k\ne 1} \left( [ g_V(x_k)+
g_S(x_k)]{\cal T}_{k1} + g_S(x_k){\cal S}_{k1}\right),
\end{equation}
where $g_V(x)=\sqrt{x}\{1-(1+x) {\rm ln}[(1+x)/x]\}$,
$g_S(x)=\sqrt{x_k}/(1-x_k) $ with $x_k=M_{N_k}^2/M_{N_1}^2$ for
$k\ne 1$,
\begin{equation}
  \label{eq:vv}
%{\cal I}_{k1}={\sum_j{\rm Im}[(Y_D^*)_{1j}\lambda_1^* \lambda_k
{\cal T}_{k1}={{\rm Im}[(Y_D Y_D^\dagger)_{k1}^2]
 \over (Y_D^\dagger Y_D)_{11} +|Y_{\chi_1}|^2}
\end{equation}
and
\begin{equation}
%{\cal J}_{k1}={\sum_j{\rm Im}[(Y_D^*)_{1j}\lambda_1
{\cal S}_{k1}={{\rm Im}[(Y_D Y_D^\dagger)_{k1}(Y_{\chi_k}
Y_{\chi_1}^\dagger)]
 \over (Y_D^\dagger Y_D)_{11} +|Y_{\chi_1}|^2}.
\end{equation}
%Notice that the term proportional to ${\cal S}_{k1}$ comes from the
% interference of the tree-level diagram with diagram (d).

As shown in \cite{skk1},  the new contributions to $\epsilon_1$ could be important for the case of  $M_{N_1}\simeq M_{N_2} \ll M_{N_3}$ for which the asymmetry %$\epsilon_1$
is approximately given by
 \begin{eqnarray}
 \epsilon_1 & \simeq & -\frac{1}{16\pi}
           \frac{M_{N_2}}{v^2}\left[\frac{Im[(Y^{\ast}_D m_{\nu}Y^{\dagger}_D)_{11}]}
           {(Y_D^\dagger Y_D)_{11}+|Y_{\chi_1}|^2}
%           \right.
%           \nonumber \\
%    & & \left.  
        +\frac{ Im[(Y_DY^{\dagger}_D)_{21}(Y_{\chi_2}Y_{\chi_1}^{\dagger})]}
                   {(Y_D^\dagger Y_D)_{11}+|Y_{\chi}|^2}\right]R~,
                   \label{epsilon2}
 \end{eqnarray}
where $R$ is a resonance factor defined by $R \equiv
|M_{N_1}|/(|M_{N_2}|-|M_{N_1}|)$, and the contributions associated with $N_3$ are suppressed. For successful leptogenesis, the
size of the denominator of $\epsilon_1$ should be constrained by the
out-of-equilibrium condition, $\Gamma_{N_1} < H|_{T=M_{N_1}}$ with
the Hubble expansion rate $H$, from which the corresponding upper
bound on the couplings $Y_{\chi_1}$ reads
\begin{eqnarray}
\sqrt{\sum_i|Y_{\chi_1}|^2}<3\times
10^{-4}\sqrt{M_{N_1}/10^9(\mbox{GeV})}.
\end{eqnarray}
%%%%%
However, the size of $Y_{\chi_2}$ is not constrained by the
out-of-equilibrium condition, so large value of $Y_{\chi_2}$ is allowed
for which the second term of Eq. (\ref{epsilon2}) can dominate over
the first one and thus the size of $\epsilon_1$ can be enhanced.
%%%%
Taking  $Y_{\chi_2}=\kappa_i
(Y_D)_{2i}$ with constant $\kappa_i$ and $(Y_D)_{1i}\sim Y_{\chi_1}$, the upper limit of
the second term of Eq. (\ref{epsilon2}) is given in terms of
$\kappa$ by $\kappa_i M_{N_2}\sqrt{\Delta m_{atm}^2}R/16\pi v^2$.

The generated B-L asymmetry is given by  $Y^{SM}_{B-L}=-\eta
\epsilon_1 Y^{eq}_{N_1}$, where $\eta$ is the efficient factor and 
$Y^{eq}_{N_1}$ is the number density
of the right-handed heavy neutrino in thermal
equilibrium at high temperature given by $Y^{eq}_{N_1}\simeq
\frac{45}{\pi^4}\frac{\zeta(3)}{g_{\ast}k_B} \frac{3}{4}$ with
Boltzmann constant $k_B$ and the effective number of degree of
freedom $g_{\ast}$. In this model, the new process of type $\chi \eta \rightarrow lH^0$
generated through virtual $N_{2(3)}$ exchange will dominantly
contribute to $\eta$.
To a good approximation, the efficiency factor can be easily estimated by replacing
$M_{N_1}$ in the case of the typical seesaw model with $M_{N_1} (Y_{\chi_2}/(Y_D)_{2i})^2$ \cite{Buchmuller}.
Then, successful leptogenesis  can be achieved
for $M_{N_1}\sim 10^{4}$ GeV, provided that
$\kappa_i=Y_{\chi_2}/(Y_D)^{\ast}_{2i}\sim 10^{3}$ and $\delta M_N (\equiv M_{N_2}-
M_{N_1})\sim O (\mbox{GeV})$,
%$M_{N_2}/M_{_1}\sim 10^3$,
which alleviate the severe fine-tuning for the mass difference between
$M_{N_1}$ and $M_{N_2}$ required for resonant leptogenesis \cite{resonant-lepto}.
Note  that the choice of parameters as above is similar to the case of
the typical seesaw model with $M_{N_1}\sim 10^4$ GeV and the effective neutrino mass
$\tilde{m}_1\sim 10^{-3}$ eV for successful leptogenesis \cite{Buchmuller}.

\vspace{0.3cm}
\section{Production of Dark Matter} \label{dmp}
Now, let us investigate how the relic abundance of the singlet neutrino, $\chi$, as a keV dark matter
can be achieved.
Since we assume that the Yukawa couplings $Y_{\chi_1}$ is very small so as to achieve small neutrino masses as well as tiny mixing angle
between sterile neutrino and active neutrino, the interactions with the couplings $Y_{\chi_1}$ were never in
thermal equilibrium in early Universe.
As shown before, in order to achieve low scale leptogenesis, large values of $Y_{\chi_2}$ is essential, so
the interactions with $Y_{\chi_2}$ may play an crucial role in achieving right amount of relic abundance.
In this work, we will show that the relic abundance of $\chi$ can be achieved via so-called freeze-in process through out of equilibrium decay of  the scalar $\eta$ \cite{freeze-in 1}.

The decays of $\eta$ into a pair of $\chi$ can occur  via the interaction term generated after heavy Majorana neutrino $N_{2(3)}$ decoupled and $\eta$ got vacuum expectation value, $|Y_{\chi_{2}}|^2 (v_{\eta} /M_{N_2}) \eta \bar{\chi}\chi$.
We consider the possibility that the scalar  interactions of $\eta$ with SM Higgs $H$ keep $\eta$ in equilibrium down to a rather low temperature of $m_{\eta}$, so that $\eta$ remains in thermal equilibrium and behaves as a bath particle above that.
The decay width of $\eta \rightarrow \chi \chi $ is given by
\begin{eqnarray}
\Gamma(\eta\rightarrow \chi \chi) \simeq \frac{|Y_{\chi_2}|^4}{32\pi}\frac{v_{\eta}^2}{M_{N_2}^2}m_{\eta}. \label{decay-eta}
\end{eqnarray}
That decay process happens out of equilibrium at $T<m_{\eta}$. Then the abundance of $\chi$ becomes
so flat that it can give rise to right amount of the relic density.
The relic density of $\chi$ is obtained by solving the Boltzmann equation for $n_{\chi}$ \cite{freeze-in 1},
\be
\frac{dn_{\chi}}{dt}+3 H n_{\chi} = \int d\Pi_{\eta} d\Pi_{2} d\Pi_{3}(2\pi)^4\delta^{(4)}(p_{\eta}-p_{2}-p_{3})
\sum_{spin}|M|^2_{\eta\rightarrow \chi \chi} f_{\eta},
\ee
where $p_2$ and $p_3$ are the momentum of the outgoing particles,  $d\Pi_{i}=d^3p_i/(2\pi)^3 2 E_i$ and the initial abundance of $\chi$ is assumed to be negligible so that we can set $f_{\chi}=0$.
Using the definition of the partial decay width for $\eta \rightarrow \chi \chi$, $\Gamma_{\eta}$, given in Eq.(\ref{decay-eta}), we can rewrite the above equation as follows;
\be
\frac{dn_{\chi}}{dt}+3 H n_{\chi} &=& 2 \int \frac{d^3p_{\eta}}{(2\pi)^3} \frac{\Gamma_{\eta} f_{\eta} m_{\eta}}{E_{\eta}} \nonumber \\
&=&  \frac{\Gamma_{\eta} m_{\eta}}{\pi^2}\int dE_{\eta} \sqrt{E^2_{\eta}-m^2_{\eta}}e^{-E_{\eta}/T} \nonumber \\
&=& \frac{\Gamma_{\eta}m_{\eta}^2}{\pi^2} T K_1(m_{\eta}/T) ,
\ee
where $K_1$ is the first modified Bessel function of second kind. Here we have used the Maxwell-Boltzmann approximation $f_{\eta}\simeq e^{-E_{\eta}/T}$.
Rewriting in terms of the yield, $Y_{\chi}=n_{\chi}/S$,
\be
\frac{dY_{\chi}}{dT}=-\frac{\Gamma_{\eta}m_{\eta}^2}{\pi^2} \frac{ K_1(m_{\eta}/T)}{SH},
\ee
where $S=2\pi^2 g^S_{\ast}T^3/45$ and $H=1.66 T^2 \sqrt{g^{\rho}_{\ast}}/M_{Pl}$.
Using $x=m_{\eta}/T$,
\be
Y_{\chi}\simeq \frac{45}{(1.66)2\pi^4}\frac{M_{Pl}\Gamma_{\eta}}{m^2_{\eta}g^S_{\ast}\sqrt{g^{\rho}_{\ast}}}\int^{x_{max}}_{x_{\min}}K_1(x) x^3 dx
\ee
Taking $x_{max}=\infty $ and $x_{min}=0$, we can get
\be
Y_{\chi}\simeq \frac{135}{4\pi^3 (1.66) g^S_{\ast}\sqrt{g^{\rho}_{\ast}}}\left(\frac{M_{Pl}\Gamma_{\eta}}{m^2_{\eta}}\right)
\ee
\be
\Omega_{\chi} h^2 
% \simeq \frac{1.09\times 10^{27}}{g_{\ast}^S \sqrt{g_{\ast}^{\rho}}}
%\frac{m_{\chi} \Gamma_{\eta}}{m_{\eta}^2} \nonumber \\
 \simeq \frac{1.09\times 10^{27} }{16 \pi g_{\ast}^S \sqrt{g_{\ast}^{\rho}}}
\frac{m_{\chi} }{m_{\eta}}\frac{v^2_{\eta}}{M^2_{N_2}}|Y_{\chi_2}|^4
\ee

For a decoupling temperature of around 70 GeV, as is the case here, we can easily take $g_{\ast}^S=g_{\ast}^{\rho}$ to a very good approximation. Calculating the number of degrees of freedom we get $g_*^{(S,\rho)}=86.25$ in this case. Thus, the formula for the relic density can easily give us the desired value of $\Omega_{\chi} h^2=0.1198$ \cite{planck} for our preferred choice of parameters given as $v_\eta=100$ GeV, $m_\eta=71.3$ GeV, $Y_{\chi_2}=10^{-3}$, $M_{N_2}=15$ TeV and $m_\chi=7.1$ keV. This shows that we can easily generate the required lepton asymmetry and the correct relic density with a relatively low seesaw scale in this model.

We would like to further investigate whether the sterile neutrino dark matter in this model behaves as a warm or cold dark matter candidate. A knowledge of its free-streaming length, though does not provide the entire picture, can give us some indication towards our goal. The free streaming length depends not only on its mass but also on the production mechanism of the dark matter particle \cite{kusenko}. The frozen-in production mechanism at $T_{prod} \sim 100~$GeV leads to a much colder dark matter candidate compared to typical warm dark matter production via Dodelson-Widrow mechanism \cite{dw}. This is also true for our model where the sterile neutrino dark matter candidate has a shorter free streaming length than typical warm dark matter but still is much larger than that of cold dark matter. A detailed analysis of the free-streaming length or the  determination of the velocity profile for the dark matter is beyond the scope of this work. 

\vspace{0.3cm}
\section{Dark matter decay}

The $Z_2$ symmetry introduced in the model is broken when the scalar field $\eta$ gets a vacuum expectation value. As a result the dark matter candidate is no longer stable and can decay. The only possible decay channels for the scalar dark matter $\chi$ would be into three active neutrinos or into an active neutrino along with the emission of a photon \cite{pal}.

\begin{figure}[h!]
\begin{center}
\includegraphics[width=3.2in]{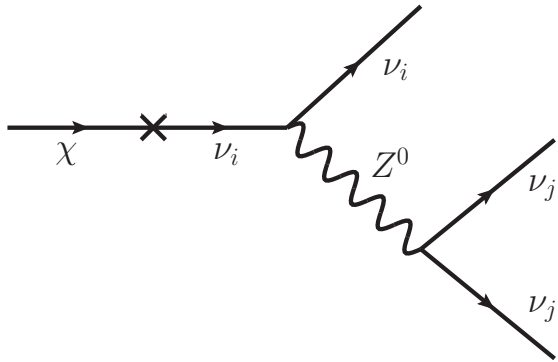}
\caption{Sterile neutrino decay into three neutrinos}
\label{fig:3nu}
\end{center}
\end{figure}

The primary decay channel for the final state with three active neutrinos is shown in Fig.~{\ref{fig:3nu}}. The decay width for this process is given as
\begin{eqnarray}
\Gamma_{3 \nu} &\simeq & \sin^2 {2\theta_{\chi}}~G_F^2 \left( \frac{m_\chi^5}{768 \pi^3}\right) \notag \\
&\simeq &8.7 \times 10^{-31}~\text{sec}^{-1}\left(\frac{\sin^2 {2\theta_\chi}}{10^{-10}}\right) \left(\frac{m_\chi}{1~\text{keV}}\right)^5
\end{eqnarray}
where $G_F \approx 1.166 \times 10^{-11}$ $\text{MeV}^{-2}$ is the Fermi constant, $m_\chi$ is the mass of the dark matter candidate and $\theta_\chi$ is the mixing angle between the singlet $\chi$ and the active neutrinos.

\begin{figure}[h!]
\begin{center}
\includegraphics[width=6.0in]{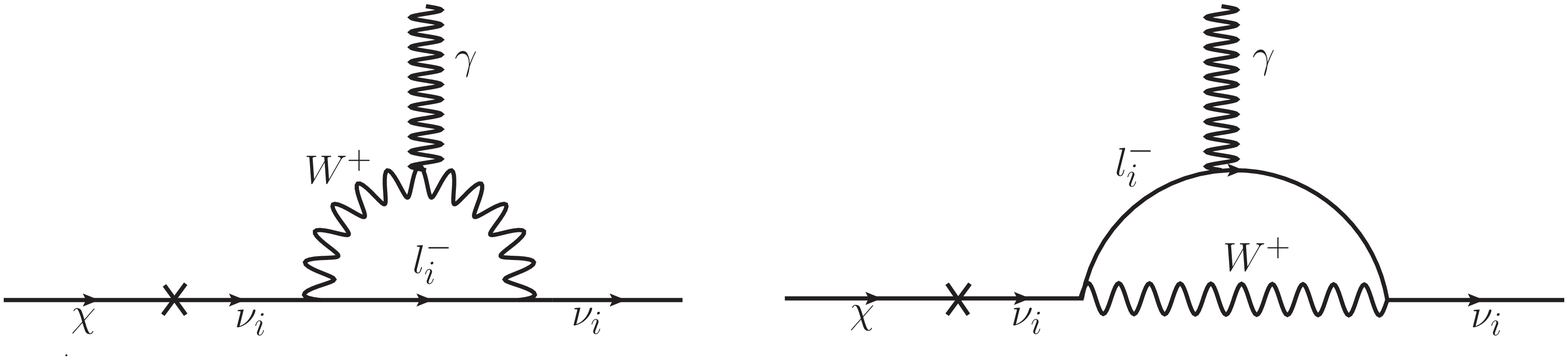}
\includegraphics[width=6.0in]{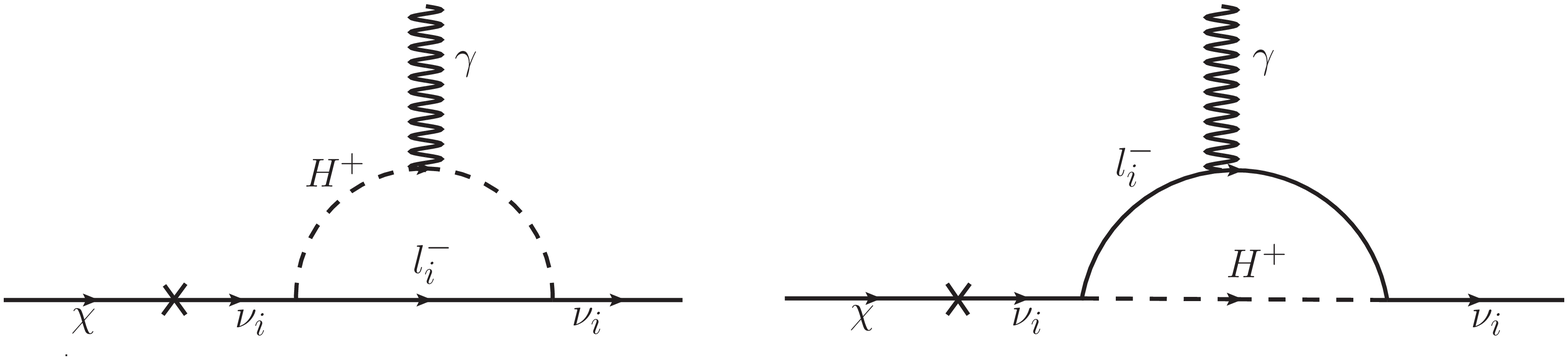}
\caption{Radiative decay of sterile neutrino dark matter}
\label{fig:nugamma}
\end{center}
\end{figure}

The processes for the radiative decay of the dark matter is shown in Fig.~{\ref{fig:nugamma}}.
In these processes, the final state is a neutrino accompanied by the radiative emission of a photon with an energy of $\frac{m_\chi}{2}$. If the mass of the dark matter particle is 7.1 keV, the emitted photon will have an energy of 3.55 keV. This can easily explain the recent observation of a 3.55 keV X-ray line signal detected in the Andromeda galaxy and many other galaxies including the Perseus galaxy. 

The decay width for these processes are inversely proportional to the mass of the charged lepton inside the loop and hence an electron would produce a much stronger limit on the lifetime of the dark matter particle. The smallness of the electron Yukawa coupling implicates that the contribution from the first pair of diagrams which involve the gauge interactions far exceed the contribution form the other pair involving Yukawa interactions. Considering the masses of the $W$ boson and the Higgs boson being of the same order the second pair of diagrams are suppressed by a factor of $\left( \frac{Y_e}{g} \right)^2$ $\sim$ $10^{-11}$, where $Y_e$ is the electron Yukawa coupling and $g$ is the $SU(2)_L$ gauge coupling. Hence the first pair of diagrams are the ones which provide any meaningful limit on the dark matter lifetime for its radiative decay.

The combined decay width for the radiative decay processes involving the gauge couplings is given as 
\begin{equation}
\Gamma_{\nu \gamma} \simeq 6.8\times10^{-33}~\text{sec}^{-1} \left(\frac{\sin^2 {2\theta_\chi}}{10^{-10}}\right) \left( \frac{m_\chi}{1~\text{keV}}\right)^5
\end{equation} 

Thus one can see that for $m_\chi$= 7.1 keV and $\sin^2 {2\theta_\chi}\sim 10^{-10}$ the decay width will be $\Gamma_{total}\sim~10^{-26}~\text{s}^{-1}$ which leads to a lifetime much larger than the age of the universe. Hence the dark matter can be considered to be relatively stable in the time frame of our universe's age.

\section{Conclusion}

In this work we consider a simple extension of the Standard Model with an extra singlet scalar ($\eta$), a singlet fermion ($\chi$) and heavy right-handed neutrinos ($N_i$). This model can consistently explain the light neutrino masses along with the baryon asymmetry of the universe. The baryon asymmetry is generated through leptogenesis, the lepton asymmetry being generated by the decay of a heavy neutrino. A new contribution corresponding to the self energy correction of the heavy neutrino helps in generating the required lepton asymmetry at low scale. The Yukawa interactions between singlet neutrino $\chi$ and $N_2$ can play an important role in connecting this low scale leptogenesis with the relic density of the keV dark matter candidate. In this model, the singlet fermion $\chi$ can be identified as a dark matter candidate of mass 7.1 keV. It decays with a lifetime much larger than the age of the universe, producing a final state photon with $E_\gamma=\frac{m_\chi}{2}$. This explains the recently observed 3.55 keV photon signal from the galactic center. We present a benchmark point of the unknown parameter set for which the required lepton asymmetry and relic density of the keV dark matter can be obtained.

Thus we have presented here a simple extension of the Standard Model which can connect the smallness of the neutrino mass, baryon asymmetry of the universe and the recently observed 3.55 keV photon signal under a single framework. The model also provides a good dark matter candidate with the correct relic density.  

\vspace{0.4in}
%\newpage

\noindent {\bf Acknowledgment:}
This work is supported in part by the NRF
grant funded by Korea government of the MEST (No.2014R1A1A2057665).
\\

\end{document}